\documentclass[%
twocolumn,
a4paper,
10pt,
superscriptaddress,
showpacs,preprintnumbers,
nobibnotes,
 amsmath,amssymb,
 aps,
prl,
letterpaper,
]{revtex4-1}

\usepackage{url}
\usepackage{graphicx,rotating,amsmath,color,dsfont,natbib}

\usepackage[active]{srcltx}

\newcommand{\be}{\begin{equation}}
\newcommand{\ee}{\end{equation}}
\newcommand{\bea}{\begin{eqnarray}}
\newcommand{\eea}{\end{eqnarray}}
\newcommand{\up}{\uparrow}
\newcommand{\down}{\downarrow}
\newcommand{\bwt}{\begin{widetext}}
\newcommand{\ewt}{\end{widetext}}
\newcommand{\ham}{\mathcal{H}}

\newcommand{\ra}{\rangle}
\newcommand{\la}{\langle}
\newcommand{\bsb}{\begin{subarray}}
\newcommand{\esb}{\end{subarray}}
\newcommand{\largem}{\!\!}

\newcommand{\gb}{\bar{G}}
\newcommand{\tw}{\tilde{\omega}}

\newcommand{\vecv}[2]{
\left(\largem
 \begin{tabular}{c}
  $#1$ \\
  $#2$
  \end{tabular}
  \largem
\right)
}

\newcommand{\vech}[2]{
\left(\largem
 \begin{tabular}{c}
  $#1$  $#2$
  \end{tabular}
  \largem
\right)
}

\newcommand{\mat}[4]{
\left(
\largem
 \begin{tabular}{cc}
  $#1$ & $#2$ \\
  $#3$ & $#4$
  \end{tabular}
  \largem
\right)
}

\begin{document}
\title{Chebyshev-BdG: an efficient numerical approach to inhomogeneous superconductivity}

\author{L. Covaci}
\affiliation{
Department Fysica, Universiteit Antwerpen,
Groenenborgerlaan 171, B-2020 Antwerpen, Belgium}
\affiliation{
Department of Physics and Astronomy, University of
  British Columbia, Vancouver, British Columbia, Canada V6T 1Z1}
\author{F. M. Peeters}
\affiliation{
Department Fysica, Universiteit Antwerpen,
Groenenborgerlaan 171, B-2020 Antwerpen, Belgium}
\author{M. Berciu}
\affiliation{
Department of Physics and Astronomy, University of
  British Columbia, Vancouver, British Columbia, Canada V6T 1Z1}

\begin{abstract}
We propose a highly efficient numerical method to describe inhomogeneous
superconductivity by using the kernel polynomial method in order to calculate
the Green's functions of a superconductor. Broken translational invariance of any type (impurities, surfaces or magnetic fields) can be easily incorporated. We show that limitations due to system size can be easily circumvented and therefore this method opens the way for the study of scenarios and/or geometries that were unaccessible before. The proposed method is highly efficient and amenable to large scale parallel computation. Although we only use it in the context of superconductivity, it is applicable to other inhomogeneous mean-field theories.
\end{abstract}

\pacs{74.45.+c, 74.20.-z, 74.62.En}

\maketitle

In the past decades the mean-field description of inhomogeneous
superconductivity through the Bogoliubov-de Gennes (BdG) equations has
been highly successful in uncovering novel phenomena. Because in the
presence of broken translational invariance one needs to use a
real space formulation, the numerical simulation becomes
computationally involved. While alternative approaches to
inhomogeneous superconductivity like quasiclassical approximations or
Ginzburg-Landau methods exist, the need for a fully quantum mechanical
approach has become imperative. This is manifest in questions
regarding high-Tc superconductors for which the superconducting
coherence length is of the order of the Fermi wave length, or in
questions regarding nanoscale superconductivity for which the
superconducting coherence length is comparable to the system size.

The BdG equations have been extensively used in a multitude of
situations where translational symmetry is broken.  Examples
include the description of quasiparticles in s-wave or d-wave vortices
\cite{franz1998,hayashi1998,atkinson2009}, self-consistent calculation
of order parameters (OP) and local density of states (LDOS) near
surfaces and interfaces
\cite{tanaka2003,black-schaffer2008,covaci2006,
zhu2000,halterman2001,martin1998}, self-consistent description of
magnetic and non-magnetic impurities in superconductors
\cite{atkinson2005,valdez-balderas2006,nunner2005}, calculation of DC
Josephson currents through weak links
\cite{black-schaffer2008,covaci2006}, uncovering of the effect of
electron confinement on superconductivity
\cite{shanenko2006,*shanenko2007,*shanenko2006-1}, etc.

Throughout these studies several methods of solving the BdG equations have been
employed. First, after a discretization of the mean-field Hamiltonian one can
use the straightforward  approach of diagonalizing exactly the
resulting Hamiltonian. 
Although exact diagonalization can in principle treat any inhomogeneous situation it
has severe limitations on the size of the discretization grid. One cure is to
recover translational symmetry either by considering surfaces and interfaces or
by considering highly symmetric geometries (cylindrical or square). This way, by
using a Fourier transformation in the direction which retains the translational
invariance, one can reduce the dimensionality of the problem: for
  each value of the 
momentum vector, we have to solve the BdG equations of dimension $d-1$, where
$d$ is the dimension of the initial problem. Another way of circumventing the
size limitations of the exact diagonalization is the use of super-cells. This
is done by considering an inhomogeneous finite size region which is then
 replicated in all directions. The super-cell method is thus able to decrease
the spacing between eigen-energies and obtain a much smoother LDOS. This is
again achieved by diagonalizing the Hamiltonian of the finite size region for
each momentum vector defined by the super-cell lattice.

A completely different approach is based on approximating the Green's
functions. In this case the eigen-energies will appear as poles of the
Green's function
while the wave-functions amplitudes will appear as weights of the poles. One
such method is the recursive method based on the Lanczos procedure
\cite{martin1998,litak1995}. The approach we use here is similar in
spirit but has several benefits when compared to the recursive method. We will
show how the Green's function can be efficiently expanded in series of
Chebyshev polynomials. The paper is organized as follows: first we will
introduce a general model Hamiltonian which is typically  used for describing
inhomogeneous superconductors. We will next present the Chebyshev-Bogoliubov-de
Gennes (CBdG) method and show, by an example, how this method can be implemented.

The Bogoliubov-de Gennes equations are mean-field coupled equations which
describe the behavior of electrons and holes in superconductors. If we consider
 second quantization and work within the Nambu spinor formalism, a general
Hamiltonian describing superconductivity can be written as follows:
\be
\label{eq:hamil1}
\ham=\sum_{\la i,j \ra} \vech{c_{i\up}^\dagger}{c_{i\down}} \hat{\ham}_{ij}
\vecv{c_{j\up}}{c_{j\down}^\dagger}
\ee
where $\hat{\ham}_{ij}$ is a $2 \times 2$ matrix:
\be
\label{eq:hamil2}
\hat{\ham}_{ij}=\mat{\epsilon_i-\mu}{\Delta_i}{\Delta_i^\star}{-\epsilon_i+\mu}
\delta_ { ij } +\mat {-t_{ij}} {\Delta_{ij}}
{\Delta_{ij}^\star}{t_{ij}^\star}(1-\delta_{ij}).
\ee
$\epsilon_i$ describes an on-site potential due to impurities,
$\mu$ is the chemical potential, 
$t_{ij}$ describes hopping between nearest neighbor sites while
$\Delta_i$($\Delta_{ij}$) are the on-site(nearest neighbor) superconducting
order parameters. The effect of a magnetic field is contained in the complex
order parameters through the usual Peierls phases
$t_{ij}=|t_{ij}|\exp(i\: \frac{\pi}{\phi_0} \int_i^jA_{ij}dl)$, where
$A_{ij}$ is the vector 
potential and $\phi_0=h/2e$ is the flux quantum. 

The quantity of interest is the $2\times2$ Green's function, which
is defined as:
\be
\label{eq:green1}
\bar{G}_{ij}(\omega)=\la vac | \vecv{c_{i\up}}{c_{i\down}^\dagger}
\hat{G}(\omega) \vech{c_{j\up}^\dagger}{c_{j\down}} |vac \ra 
\ee
where $\hat{G}(\omega+i \eta)=[\omega + i\eta-\ham]^{-1}$ and $ |vac \ra$ is the
vacuum . The diagonal and off-diagonal components are the normal and
anomalous Green's functions:
\bea
\label{eq:green2}
\bar{G}_{ij}^{11}(\omega)&=&\la c_{i\up}|\hat{G}(\omega)|c_{j\up}
^\dagger \ra \\
\bar{G}_{ij}^{12}(\omega)&=&\la
c_{i\down}^\dagger|\hat{G}(\omega)|c_{j\up }^\dagger \ra^\ast
\eea
where $|c_{i\up }^\dagger \ra=c_{i\up }^\dagger|vac \ra$ creates a
spin-up electron and $|c_{i\down } \ra=c_{i\down }|vac \ra$ destroys a
spin-down electron. For finite temperatures the expectation
value also contains a thermal average. 
 
As mentioned before, the Green's function can be approximated by using a
Lanczos procedure to invert  the Hamiltonian
\cite{martin1998,litak1995}. This method has proven to be efficient mostly
in the homogeneous case or when the Lanczos procedure can be easily
extrapolated. The need of extrapolation is of utmost importance because due to
numerical round-off errors the Lanczos procedure is unstable after a number of
iterations. Re-orthogonalization schemes exist but the method becomes less and less efficient.

We therefore propose  another approach to approximate the Green's function.
Our method is based on the Kernel Polynomial Method \cite{weisse2006} which expands
the single particle Green's function into a series of Chebyshev polynomials.
Any integrable function $f(x):[-1,1]\rightarrow \mathds{R}$ can be expanded as:
\bea
f(x)&=&\frac{2}{\sqrt{1-x^2}} \sum_{n=0}^\infty a_n T_n(x), \\
a_n&=&\frac{1}{\pi(1+\delta_{0,n})}\int_{-1}^1 f(x) T_n(x) \: dx,
\eea
where $T_n(x)=\cos[n \: \arccos(x)]$ are the Chebyshev polynomials of first kind and $\delta_{0,n}$ is the Kronecker delta function.
They are described by the following recursive relations
\be
\label{eq:rec}
T_{n+1}(x)=2xT_n(x)-T_{n-1}(x).
\ee
In order to be able to expand the Green's function, one needs first to rescale
the Hamiltonian such that its spectrum is contained in the $[-1,1]$ interval.
We therefore have to work with the rescaled Hamiltonian
$\tilde{\ham}=(\ham-\mathds{1}b)/a$ and rescaled energies
$\tilde{E}=(E-b)/a$, $\tilde{\omega}=(\omega-b)/a$ where $a=(E_{max}-E_{min})/(2-\eta)$ and
$b=(E_{max}+E_{min})/2$, where $\eta>0$ is a small number. It is not essential to have accurate bounds on the
spectrum, thus a quick Lanczos procedure to find $E_{max}$ and $E_{min}$ can be
used.

If we consider the regular Green's function in the Lehman representation we can write for its imaginary part:
\be
\label{eq:imag1}
\Im{\gb^{11}_{ij}(\omega+i\eta)}=-\pi\sum_k \la c_{i\up}| k \ra \la k | c_{j\up}^\dagger \ra \delta(\omega - E_k)
\ee
where $\{|k\rangle\}$ are the eigenvectors and $E_k$ are the eigenvalues. If we now use the Chebyshev expansion we write:
\be
\label{eq:imag2}
\Im{\gb^{11}_{ij}(\tw+i\eta)}=-\frac{2}{\sqrt{1-\tw^2}}\sum_n a^{11}_n(i,j) T_n(\tw),
\ee
where the coefficients can be calculated as the matrix elements of the Chebyshev polynomial of order $n$ of $\ham$ :
\bea
\label{eq:coeff1}
a^{11}_n(i,j)&=&\int_{-1}^1 \frac{dE}{1+\delta_{0,n}} \sum_k \la c_{i\up}| k \ra \la k | c_{j\up}^\dagger \ra \delta(\omega - E_k) T_n(E) \nonumber \\
&=& \la c_{i\up}| T_n(\ham) |c_{j\up}^\dagger \ra / (1+\delta_{0,n}).
\eea
With the use of a Kramers-Kr\"onig relation, the real part of the Green's function can now be expanded in terms of Chebyshev polynomials of second kind $U_n(x)=\sin[(n+1)\arccos(x)]/\sin[\arccos(x)]$ \cite{weisse2006}:
\be
\label{eq:real1}
\Re\gb^{11}_{ij}(\tw+i\eta)=-2\sum_{n=1}^\infty a^{11}_n(i,j) U_n(\tw)
\ee

After combining Eq.~(\ref{eq:imag2}) and Eq.~(\ref{eq:real1}), the full Green's function can be written as:
\be
\label{eq:gf1}
\gb_{ij}^{11}(\tw)=\frac{-2 i}{\sqrt{1-\omega^2}} \sum_{n=0}^\infty a_n^{11}(i,j)
e^{-in \arccos(\tw)},
\ee
with
$a^{11}_n(i,j)=\la c_{i\up}| T_n(\ham) |c_{j\up}^\dagger \ra / (1+\delta_{0,n})$.
The procedure of finding the anomalous Green's function is identical, only the coefficients will be modified accordingly:
\be
\label{eq:coeff2}
a^{12}_n(i,j)=\la c_{i\down}^\dagger| T_n(\ham) |c_{j\up}^\dagger \ra / (1+\delta_{0,n})
\ee

The most important part of the calculation has now shifted to the calculation of the expansion
coefficients $a_n^{\alpha \beta}(i,j)$. Fortunately, due to the recurrence
relation between Chebyshev polynomials, see Eq. (\ref{eq:rec}),
  these moments can be obtained efficiently through a recursive procedure.

If we define $|j_n \ra = T_n(\ham) |c_{j_\up}^\dagger \ra$, then after using the recursive property of Chebyshev polynomials [\ref{eq:rec}] we can write:
\be
\label{eq:ref2}
|j_{n+1} \ra = 2\ham |j_{n} \ra - |j_{n-1} \ra,
\ee
where $|j_0 \ra = |c_{j\up}^\dagger \ra$ and $|j_1 \ra = \ham
|c_{j\up}^\dagger \ra$. At each iteration step $a^{1\alpha}_n(i,j)=\la
\alpha | j_n \ra$, where $\la 1 | = \la c_{i\up} |$ and $ \la 2 | =
\la c_{i\down}^\dagger|$. It is important to note at this point that
in the recursion defined by Eq.~[\ref{eq:ref2}] the most intensive
computation is a sparse matrix - vector multiplication. Moreover, the
Hamiltonian matrix does not have to be stored since it always has the
same form, thus allowing for simple rules for the
multiplication. Another great benefit of this method is the
possibility of obtaining in a single iteration all the normal and anomalous
Green's functions, $\gb^{1\alpha}_{ij}(\tw)$, for all $\{i\}$ and
$\{\alpha\}$ when the starting vector is $| c_{j\up}^\dagger \ra$. As
explained in Ref.~\onlinecite{weisse2006}, because we can only keep a
finite number of terms in the expansion, one needs to convolute the
approximated function with kernel polynomials in order to remedy the
effect of Gibbs oscillations. This is imperative when approximating
Green's functions because of their discontinuous nature; the imaginary
part is a summation over delta functions. We will use the Lorentz
kernel \cite{weisse2006}, since it allows for the manipulation of a
Lorentzian broadened delta function. The expansion has the
same form, but the coefficients have to be multiplied by  factors
defined by the Lorentz kernel: 
\be
\tilde{a}^{\alpha \beta}_n(i,j)=a^{\alpha \beta}_n(i,j)\frac{\sinh[\lambda(1-\frac{n}{N})]}{\sinh(\lambda)},
\ee
where $N$ is the total number of terms in the expansion and $\lambda$ is a real number. If we write the Lorentzian approximation  as $\delta(x)=1/\pi\lim_{\epsilon \rightarrow 0 }\epsilon / (x^2+\epsilon^2)$,
then there is a direct relation between the broadening $\epsilon$ and
$\lambda$: $\epsilon=\lambda/N$. This allows for a good control over
the broadening of the Green's function's features, whether used artificially at
zero temperature or naturally at finite temperature. As we will show
later, in certain situations where interference between parts of the
considered system is important, one needs a large number of
coefficients in order to accurately obtain the Green's function. In
that case the only way to keep the broadening constant is by changing
$\lambda$ accordingly. 

Once the Green's functions are known, it is straightforward to calculated physically relevant quantities. The local density of states can be calculated as:
\be
\label{eq:ldos}
N^{\up(\down)}(E,i)=-\frac{1}{\pi} \Im{\gb^{11(22)}_{ii}}(E).
\ee
The electron density is:
\be
n_{i}=\int_{-\infty}^\infty \left[ N^\up(E,i) + N^\down(E,i) \right] f(E) dE.
\ee
The order parameter, $\Delta_{ij}= U_{ij}\la c_{i\up} c_{j\down} \ra $ is:
\be
\label{eq:op1}
\Delta_{ij}=iU_{ij} \int_{-E_c}^{E_c} \gb^{12}_{ij}(E) (1-2f(E)) dE,
\ee
where $E_c$ is a cutoff energy (Debye energy for conventional superconductors or the bandwidth for cuprates).
The current density between grid points $i$ and $j$ is:
\be
J_{ij}^{\up(\down)}=\frac{-1}{\pi}\int \Im[i\: t_{ij} \gb^{11(22)}_{ij}(E) - i\: t_{ij}^\star \gb^{11(22)\star}_{ij}(E) ] f(E) dE.
\ee

One of the great benefits of this method is that the Green's function is calculated separately for each grid point thus allowing for a trivial parallel implementation. An iteration can be started on a separate CPU for each grid point with a given order parameter profile. Next the order parameter for that grid point is updated in the Hamiltonian in order to achieve self-consistency. 
The method is general and it can be applied not only to any mean-field Hamiltonian but also to more complex band structures, multi-band superconductivity and even to three dimensional systems. Of course the number of operations increases dramatically but the calculation can be done even on a desktop computer since the Hamiltonian is sparse.

\begin{center}
 \begin{figure}[t]
  \includegraphics[angle=-90,width=\columnwidth]{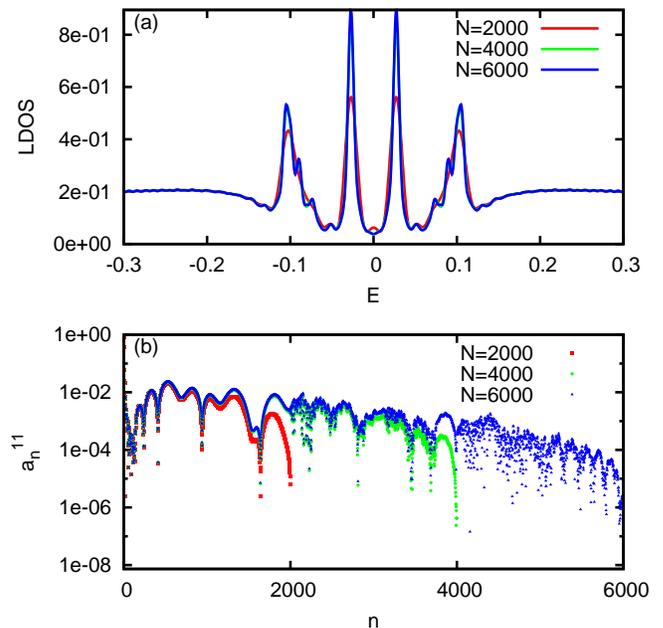}
 
  \caption{LDOS and $a_n^{11}$ at the surface of a planar s-wave superconductor/normal metal system.  $L_x^S/a=300$, $L_x^N/a=20$ and $L_y/a=500$.}
 \label{fig1}
 \end{figure}

\end{center}

\begin{center}
 \begin{figure}[ht]
  \includegraphics[angle=-90,width=\columnwidth]{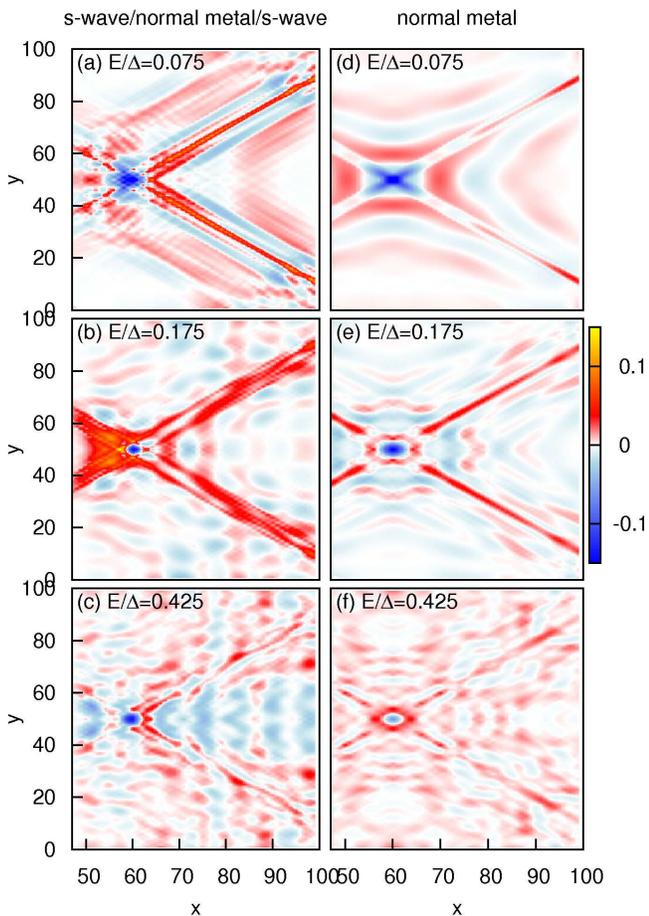}
  \caption{Local density of states around an impurity in a normal
  metal for various energies. For panels (a)-(c) the normal metal is
  sandwiched between two s-wave superconducting regions  located
  at $x<50a$ and $x>100a$ respectively, while for
  panels (d)-(f) the whole system is normal. The LDOS for the
   corresponding clean systems (no impurity) is subtracted in
  both cases, for clarity.  } 
\label{fig2}
 \end{figure} 
\end{center}

As an example we will show how the LDOS depends on the number of Chebyshev
coefficients. We consider first a planar system
composed of a normal metal of length $L_x^N=380a$ and an s-wave
superconductor of length $L_x^S=20a$ while $L_y=500a$. In
Fig.~\ref{fig1}(a) we plot the LDOS at the surface of the normal
metal region. Choosing $\Delta=0.1t$ such that $\xi \approx 7a$, we
observe in the LDOS the appearance of Andreev bound states below the
superconducting gap. In Fig.~\ref{fig1}(b) we plot the moments of the
Chebyshev expansion for three iteration sequences. Here we choose a
constant broadening $\epsilon=0.001t$, thus the coefficient
$\lambda=\epsilon/N$ will modify the Chebyshev moments for each
sequence. We observe an oscillatory behavior of the Chebyshev moments
which is given by the interference of quasiparticles scattering off
the normal/superconducting  and normal/vacuum interfaces. Note that the Chebyshev iteration
is equivalent to a propagation of a quasiparticle defined by the
starting vector $|i\rangle$. Interestingly the LDOS is not converged
within $\epsilon$ for $N=2000$, instead  a larger number of
moments is needed. It is exactly for these type of systems that a stable method
is essential. When interference between quasiparticles scattered of distant regions of the system is important, an accurate solution requires a large number of moments. The recursion
method based on the Lanczos method fails in these situations. 

To illustrate the power of the method we show in Fig.~\ref{fig2} the
LDOS for a s-wave SC/normal metal/s-wave SC of size
$50a/50a/50a \times 100a$ in the presence of a non-magnetic impurity
in the normal region $V_i=V\exp[-(\mathbf{r}_i-\mathbf{r}_{i0})^2/a]$ with $V=2t$ and
$r_{i0}=(60a,50a)$. The left panels show the LDOS around the impurity
for various sub-gap energies while the right panels show the LDOS for
a homogeneous normal system. Modifications of the LDOS induced by the
impurity are seen in both cases but for the s-wave/normal metal/s-wave system extra states
are induced by the interference of quasiparticles undergoing Andreev
reflection at the superconductor/normal metal interface and specular reflection at the
impurity site. Andreev states of the clean multilayer system are destroyed by
the impurity but new states appear due to impurity scattering.  

In conclusion we have introduced and demonstrated a new method of solving the
mean-field self-consistent BdG equations by expanding the Nambu
Green's functions in terms of Chebyshev polynomials. Because the
method is stable the results are arbitrarily accurate since the
accuracy is given by the number of moments kept in the expansion. The
most expensive numerical operation is a sparse matrix - vector
multiplication, thus allowing for large sized systems to be solved
with little memory requirements. Moreover, since each grid point is
calculated separately the method is amenable to trivial parallel
implementations. The present method can be easily expanded to consider
complex band structures, multi-band superconductivity, three
dimensional system and other mean-field Hamiltonians. 

Acknowledgments:  This work was supported by the Flemish Science
Foundation (FWO-Vl), CIfAR and NSERC. Discussions with Frank Marsiglio are gratefully acknowledged.

\bibliography{biblio_}

\end{document}